\documentclass[%
 aip,
 amsmath,amssymb,
 reprint,%
]{revtex4-1}

\usepackage{graphicx}% Include figure files
\usepackage{dcolumn}% Align table columns on decimal point
\usepackage{bm}% bold math
\usepackage[utf8]{inputenc}
\usepackage[T1]{fontenc}
\usepackage{mathptmx}
\usepackage{etoolbox}

\makeatletter
\def\@email#1#2{%
 \endgroup
 \patchcmd{\titleblock@produce}
  {\frontmatter@RRAPformat}
  {\frontmatter@RRAPformat{\produce@RRAP{*#1\href{mailto:#2}{#2}}}\frontmatter@RRAPformat}
  {}{}
}%
\makeatother

\begin{document}

\preprint{AIP/123-QED}

\title[DNA KMMD]{Quantum Machine Learning Corrects Classical Force Fields: Stretching DNA Base Pairs in Explicit Solvent}

% Force line breaks with \\
\author{Joshua T. Berryman$^*$}%
 \email{josh.berryman@uni.lu}
\author{Amirhossein Taghavi}
\author{Florian Mazur}
 \altaffiliation[Also at: ]{Laboratoire de Physique et Chimie Th\'eoriques, Universit\'e de Lorraine, Faculté des Sciences et Technologies,
Boulevard des Aiguillettes, B.P. 70239, F-54506 Vandoeuvre-les-Nancy, France.}
\author{Alexandre Tkatchenko}%
\affiliation{Department of Physics and Materials Science, University of Luxembourg, L-1511 Luxembourg City, Luxembourg.}

\date{\today}% It is always \today, today,

\begin{abstract}
In order to improve the accuracy of molecular dynamics simulations, classical force fields are supplemented with a kernel-based machine learning method trained on quantum-mechanical fragment energies. As an example application, a potential-energy surface is generalised for a small DNA duplex, taking into account explicit solvation and long-range electron exchange--correlation effects. A long-standing problem in molecular science is that experimental studies of structural and thermodynamic behaviour of DNA under tension are not well confirmed by simulation; study of the potential energy versus extension taking into account a novel correction shows that leading classical DNA models have excessive stiffness with respect to stretching. This discrepancy is found to be common across multiple forcefields. The quantum correction is in qualitative agreement to the experimental thermodynamics for larger DNA double helices, providing a candidate explanation for the general and long-standing discrepancy between single molecule stretching experiments and classical calculations of DNA stretching. The new dataset of quantum calculations should facilitate multiple types of nucleic acid simulation, and the associated Kernel Modified Molecular Dynamics method (KMMD) is applicable to biomolecular simulations in general. KMMD is made available as part of the AMBER22 simulation software.
\end{abstract}

\maketitle

\textbf{The definitive version of this manuscript is the publisher postprint which is open access at J Chem Phys:} \url{https://doi.org/10.1063/5.0094727} .

\section{\label{sec:Intro}Introduction}

The dominant long-range interactions in molecular systems can be treated classically with great efficiency as a pairwise Coulomb term ($U_c \propto 1/r$) plus a pairwise fluctuating dipole-dipole attraction (a dispersion interaction, $U_{d} \propto 1/r^6$), with atoms represented as spheres having fitted non-integer charge and attraction parameters, both of which in theory can be derived from \textit{ab initio} quantum mechanics. \cite{Dupradeau2010,Bleiziffer2018,Tkatchenko2009} Limitations to the accuracy of this approach are especially evident when atomic polarisability is anisotropic, or when the atoms are part of a flexible molecule in which the atomic polarisability and the partial charge can couple to the molecular conformation. These two limitations apply very much to nucleic acids where the charged backbone, anisotropically delocalised electrons in the aromatic bases, polar solvent, and close localisation of cations give a strongly many-body character to the interatomic non-bonded forces. \cite{Sponer1996,Sponer2006}\\

Although the limitations of classical forcefields are understood, with large discrepancies noted in less-standard conformations such as single stranded DNA \cite{Nganou2016}, due to the complexity of nucleic acid molecules and the important role of dispersion interactions, it has been quite difficult to arrive at solid quantum benchmark calculations from which to improve these models. Substantial progress has been made in recent years by calculation of dispersion interactions in an explicit many-body manner, working with tensors of directional atom-centred polarisabilities (for example, in the many-body dispersion method\cite{Tkatchenko2012}).  This approach was found to be quantitative for DNA stacking energies \cite{DiStasio2012} and benchmarked favourably against other methods over a dataset of `challenging' noncovalent complexes including DNA bases.\cite{Hamdani2019}\\

Evaluation of potential energies of static molecular conformations has limited relevance to the highly dynamic finite-temperature behaviour of biomolecules: ideally in order to study large nucleic acid molecules in explicit solvent and salt a potential energy surface (PES) should be defined over the nuclear coordinates. The PES must be differentiable and inexpensive to compute, scaling better than $\mathcal{O}(N^2)$ in the number of atoms. In order to integrate the accuracy of the \textit{ab initio} calculations to existing efficient infrastructure of classical forcefields, a Kernel Modified Molecular Dynamics (KMMD) approach is devised here.\\

The KMMD approach, described in detail below (\ref{sec:Methods}), uses a kernel machine to learn not the full PES, but only the part of the quantum correction to the classical PES susceptible to prediction by selected internal degrees of freedom of the nucleic acid chain.  Solvent-solvent interactions are thus calculated as usual for the chosen classical forcefield, and solvent-DNA interactions are modified only when they can be predicted by the DNA conformation. This is a form of coarse-graining: degrees of freedom which are not explicitly corrected may still be corrected on average.   DNA-DNA interactions are only modified in so far as can be predicted from the training data, for instance because information relating to bond lengths was excluded from the training data, bond lengths are only improved in the KMMD as far as they are driven by correlations with the explicitly treated torsions.  The role of solvent is significant for nucleic acids, so it is important to understand that although the KMMD correction presented is not a correction to solvent interactions, because solvent was present in the training data it is nonetheless an implicit correction to interactions in a solvated system.\\

Having implemented the KMMD method, simulations were made of a GG$\cdot$CC DNA complex (with Watson-Crick base pairing artificially imposed), under constant force at the O3' and O5' termini, before and after quantum corrections.  The observed small reduction in the work to extend with base pairing preserved was consistent with the large long-standing difference between single-molecule stretching experiments on DNA polymers and classical simulation (comparing simulations \cite{Harris2005,Roe2009} against a review of experimental data \cite{Bustamente2000}, work in experiment is 50-70\% of work in classical simulations).  While a change in work to extend single steps should imply a change of the same sign for large polymers, for the present prototype study only a small GG$\cdot$CC complex was studied, and direct experimental measurement of single-step stretching free energies in order to make a quantitative comparison is non trivial. Experimental measurement of single base stacking energy \textit{has} been made for a comparable proxy system, a nicked duplex.\cite{Yakovchuk2006} This setup is not exactly equivalent to the calculation geometry however the experimental energies are closer to the KMMD corrected calculation than to the uncorrected (\textit{vide infra}: \ref{sec:Results}).\\

Various alternative hypotheses exist in relation to the discrepancy between experimental and calculated DNA extension dynamics, and not all are mutually inconsistent.  It is possible that experimental systems contain chemical defects not accounted for in classical simulation or by KMMD: nicking of the backbone due to some form of tension-accelerated hydrolysis, for example, or occasional base-pair mismatches due to failure of polymerase enzymes.  On the simulation side, apart from the forcefield errors documented here, there is the problem of scaling up results from a simulation on a microseconds$\times$nanometers scale to experiments in the world of seconds$\times$millimeters, and from seconds$\times$millimeters to the abstraction of a fully adiabatic, quasi-equilibrium stretching. Advanced nonequilibrium data analysis using the Jarzynski equality is often applied to make the scaling-up in time, even when starting from the relatively long timescales of experiment.\cite{Liphardt2002}  A corresponding recovery of equilibrium stretching behaviour from continuous stretching simulations requires care and expertise;\cite{Park2003} in DNA there are certain very rare dynamical fluctuations such as base breathing which might well have an impact on the stretching process but which are expected to manifest only on microsecond timescales,\cite{Galindo2015} thus requiring many microseconds for effective sampling.  Finite-size thermodynamics arising through the generalisation from small to large systems are equally subtle, and difficult to treat for a complex molecule such as DNA.\cite{Manghi2008} Finally, although the present study addresses collective electronic effects in the DNA, collective electron fluctuations in the solvent are not treated beyond the nearest few molecules, and these have been shown to be important for biomolecules up to the 20\AA~scale.\cite{Stohr2019}

\section{\label{sec:Methods}Methods}

\subsection{Generation of DNA Training Data}

69 GG$\cdot$CC DNA dyad structures were sampled from existing classical atomistic simulations of 24bp (24 base pair) DNA duplexes at varying extensions documented in previous work.\cite{Taghavi2017}  Hydrogens were added to the O5' and O3' ends.  Each dyad structure was solvated with 2446 TIP3P waters \cite{Jorgensen1983} and two sodium ions (neutralising the system), and the solvent box was then allowed to equilibrate while holding the dyads in stiff restraints for 10ns. This classical simulation and processing was done using AmberTools. \cite{Case2020}  The OL15 forcefield \cite{Zgarbova2015} was used. Dynamics were classical (no nuclear quantum effects were treated) allowing all atomic masses to be set to 12 amu. 

From the initial 69 structures, fresh conformations were bootstrapped from long KMMD simulations with a total time of 1$\mu$s (sampling one per ns) to make a training set of 1000 solvated 2bp duplexes. The bootstrapping process was designed as a carousel of 69 independent runs, each restrained to have roughly constant extension O3'-O5' and each restarting after a 10ns window, to use fresh training data generated by all previous runs.  When a total of 1 microsecond dynamics was achieved (1000 frames at one per nanosecond), the bootstrapping process was stopped. 

For each training snapshot, a structure of the DNA with 20 water molecules and 2 sodium ions  surrounding it in a cluster was created. The water molecules were selected based on the 20 closest to any solute atom, this number was found to be sufficient such that backbone charges and solute-solvent hydrogen bonds were always coordinated.  Energies were found for the cluster structures using density functional theory (DFT) calculations with the hybrid PBE0 functional\cite{Perdew1996,Adamo1999,Ren2012} for semi-local electron exchange and correlation together with the MBD method \cite{Tkatchenko2012,Ambrosetti2014} for long-range electron correlation interactions (including van der Waals forces). The FHI-aims code was used to converge the electron density,\cite{Blum2009} with the associated atom-centred basis sets represented at the `intermediate' level where available (H, C, N, O) but at the `tight' level when intermediate basis sets were not available for the given element (Na, P).  Memory requirement for a neutralised dyad plus 20 water molecules was approximately 12GB, with convergence taking approximately 12 hours on 8 cores.  Two structures which did not converge within 72 hours were rejected as having unphysical geometry, leaving 998 initial reference DNA structures for use in the first stage of the machine learning calculation.  

\subsection{Kernel Method for Generalisation to PES}

The training datapoints $\left\{ \Phi_i (\vec{x}_i) \right\}$ were defined as the residuals of the quantum and classical energies: $ \Phi_i = H_Q(\vec{x}_i) - H_{AMBER}(\vec{x}_i)$.  Evaluations of $\left\{H_{AMBER}\right\}$ are made at the start of each run, for convenient re-use of the quantum data across forcefields.  Feature extraction was carried out in order to reduce the degrees of freedom of the problem, by defining a vector of dihedral angles $\vec{\phi}_i(\vec{x_i})$ based on the Cartesian coordinates.  The canonical DNA torsions  $\left(\alpha, \beta, \gamma, \delta,  \epsilon,\zeta,\chi,\nu_0,\nu_1,\nu_2,\nu_3,\nu_4 \right)$ were collected, with one occurrence per strand of the backbone torsions $\alpha, \beta, \gamma, \epsilon, \zeta$ and two occurrences per strand of the sugar torsions $\delta, \nu_{0-4}$ and the base twisting angle $\chi$, giving a total of 38 angular degrees of freedom to summarize the state of a base-pair step.  These are then (reversibly) expanded to vectors $\vec{p_i}$ of 76 Cartesian degrees of freedom by taking the $\sin$ and $\cos$ of each angular degree of freedom: $\vec{p_i}=(\sin \vec{\phi_i}, \cos \vec{\phi_i})$.\\

The feature space of transformed angles forms a toroidal manifold embedded in $\mathbb{R}^{76}$. The maximum Euclidean distance between two points is $r_{max}=2\sqrt{38}\approx12.3$ and the average distance between two random points is $\sqrt{76}\approx 8.7$.  For small angular differences, the distance in feature space is very close to the imaged distance in angular space (perturbing a single angle by 0.1 rad implies $r=0.0996$ in the feature space, 1 rad implies $r=0.959$, $\pi$ rad implies $2$). The PES correction at a feature space point $\vec{q}$ is estimated from the training set using a single-parameter Gaussian kernel:

\begin{equation}
\label{eqn:kernel}
\Phi(\vec{q}) = \frac{ \sum_i \Phi_i \exp \left( -\lVert \vec{q} - \vec{p_i} \rVert^2/\sigma^2 \right)  }{ \sum_i \exp( -\lVert \vec{q} - \vec{p_i} \rVert^2/\sigma^2 )}
\end{equation}

where $\sigma^2 := 0.1$ for all results discussed here.  In the limit of small $\sigma$ the PES becomes a Voronoi diagram in the feature space (losing differentiability), in the limit of large $\sigma$ the PES becomes flat.  For any $\sigma$ the PES deals gracefully with duplicate training points, simply averaging their contributions: this is valuable because it permits degrees of freedom not represented in the feature space to be averaged over, however the caveat is introduced that if multiple copies of training points exist, untreated degrees of freedom should preferably be sampled using the Gibbs measure at the target temperature for simulation.  The denominator in eqn.~\ref{eqn:kernel} is treated as an empirical confidence value for a given point evaluation. This kernel method, a minor variation of the extant family of kernel methods, is referred to as a `Normalised Radial Basis Function' (`nRBF') kernel machine, or equivalently as a nRBF network.\cite{Moody1989}

To calculate the force, first define $p_{i,d}$ as the displacement of point $i$ in dimension $d$ in the Cartesian feature space from the reference position of training point $i$, having $D$ treated dihedral angles. For even $d\in[0,2..2D-2]$,
$p_{i,d}=\cos\phi_{i,d/2} - \cos\phi^0_{i,d/2}$; for odd $d\in[1,3...2D-1]$, $p_{i,d}=\sin\phi_{i,(d-1)/2} - \sin\phi^0_{i,(d-1)/2}$.    $R^2_i = \sum_{d\in[0,2D-1]} p^2_{i,d}$.\\

We write a "weight" for the $i^{th}$ training point, $w_i$ as:

\[
w_i = \exp - R^2_i / \sigma^2
\]

The normalisation term follows:

\[
Z = \sum_i w_i
\]

The energy delta:

\[
\Phi = \frac { \sum_i \Phi_i w_i }{Z}
\]

Differentiating, using a primed notation for the partial derivative of a function in dimension $d$ for point $i$: $f'= \frac{\partial f}{\partial p_{i,d}}$:

\[
w' = -2 p_{i,d}  w_i / \sigma^2
\]
\[
Z' = w'
\]

Following the quotient rule:

\[
\Phi' = \frac { \Phi_i w' Z  - Z'\sum_i \Phi_i w_i }{Z^2}
\]

Substituting $Z'=w'=-2p_{i,d}w_i/\sigma^2$:

\[
\Phi' = -2p_{i,d}w_i\frac { \Phi_i Z  - \sum_i \Phi_i w_i }{\sigma^2 Z^2}
\]

Substituting $\Phi=\sum_i \Phi_i w_i / Z$:

\begin{equation}
\label{eqn:kfrc2}
\Phi' = -2 p_{i,d}  w_i\frac { \left( \Phi_i  - \Phi \right) }{\sigma^2 Z}
\end{equation}

The Cartesian feature space has dimension $2D$, where $D$ is the number of torsions studied.  Forces are not in general tangent to the subspace of the Cartesian space which can be mapped directly back to the space of angles, therefore they are projected onto the subspace.  By construction, the force gradients are linear with respect to individual training points and angles, but must be treated in pairs for the purpose of projection back to the subspace of dihedral angles. The 2D unit radius vector corresponding to angle $\phi_{i,d}$ for $d\in[1,D]$ is already known as:

\[
\hat{\vec{u}}_{i,d} = \cos\phi_{i,d}\hat{\vec{e_1}} + \sin\phi_{i,d}\hat{\vec{e_2}}
\]

\[
\hat{\vec{u}}_{i,d} = p_{i,2d}\hat{\vec{e_1}} + p_{i,2d+1}\hat{\vec{e_2}}
\]

to project orthogonal to a unit vector we subtract a term scaled by the inner product:

\begin{equation}
\vec{\Phi'_{d}} \leftarrow  \vec{\Phi'_{d}}  - (\hat{\vec{u}}_{i,d} \cdot \vec{\Phi'_{d}}) \hat{\vec{u}}_{i,d} 
\end{equation}

This vector can then be converted directly to an angular force.\\

\subsection{Validation of Machine Learning Approach}

\begin{figure}[ht]
\begin{centering}
         \includegraphics[width=1.0\columnwidth]{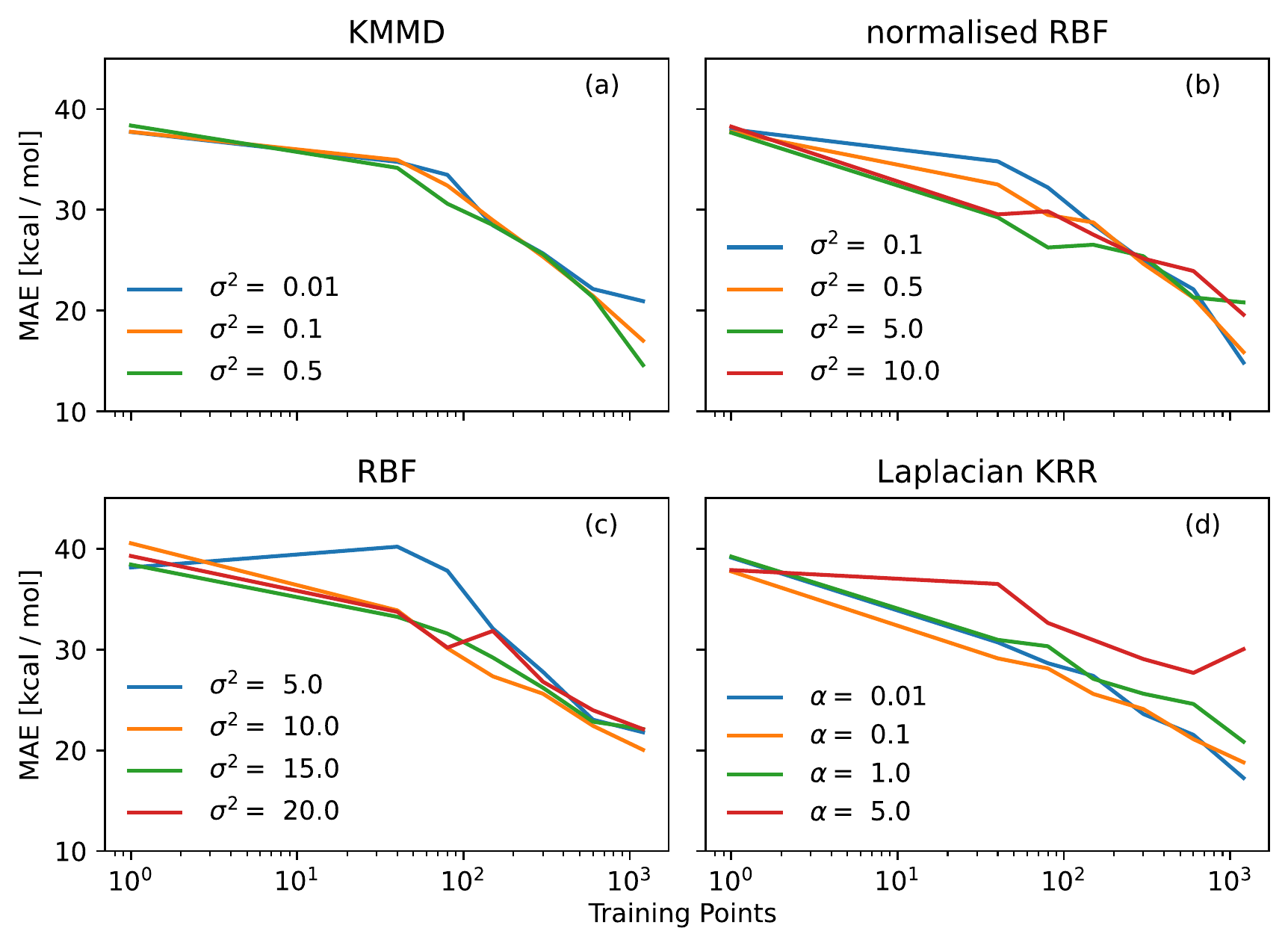}
\end{centering}
    \caption{\label{fig:learning} Validation of KMMD implementation \textbf{(a)} against related machine learning methods, including a conventional nRBF \textbf{(b)} with weights $\Phi_i$ as trained parameters.  Performance is tested against unseen validation data for increasing amounts of seen training data. The basic KMMD approach is stable with respect to parameters and performs well.} 
\end{figure}

Validation of the KMMD approach was carried out by subsampling a set of 1232 frames, then evaluating the trained system against the remaining unseen frames.  Mean Absolute Error (MAE) values shown are averages over five independent selections of training/test sets, noise is greater for larger training sets due to the smaller validation sets. The KMMD was robust at predicting energies regardless of the $\sigma^2$ parameter and made more efficient use of training data than the other, more sophisticated, approaches.  The ``normalised Radial Basis Function Network'' (nRBF) has an identical functional form to that of the KMMD (eqn.~\ref{eqn:kernel}) except in that the $\Phi_i$ are not training point energies, but arbitrary values fixed by a least-squares fit to the training data.\cite{Bugmann1998}  In theory this fitting should allow a performance improvement,
however with a relatively modest number of training points (as in this example, where the number of centres is still small enough to be computationally cheap at evaluation time), then there is no gain and potentially some minor loss due to under-constraint of the fitting problem.  Relaxing the constraint that the weights should be normalised (removing the denominator of eqn.~\ref{eqn:kernel}) converts the system to a standard RBF network, this removal of a constraint gives even greater freedom in fitting however no benefit was observed. 

Kernel Ridge Regression (KRR) is a generalisation of the RBF network which has been used successfully to fit atomisation energies of small molecules \textit{in vacuo} using a feature space constructed around the Coulomb Matrix of atom-atom distances \cite{Hansen2013}, the Laplacian kernel $e^{-|x|}$ recommended in this related example was tested in the present feature space and found to perform about as well as the Gaussian nRBF for some choices of the KRR regularisation parameter $\alpha$.  Although the KMMD performed best on the present dataset, all methods have the potential for optimisation and for efficiency gains against larger datasets. In the present examples all training points became nodes of the evaluation, for larger training sets it is possible to merge or prune training points, or even to use non-physical constructed points as RBF centres, gaining efficiency at evaluation time in exchange for a small or zero loss of accuracy. 

\subsection{Constant Force Simulations}

In order to monitor the influence of extensional force on the DNA dynamics, 36 parallel simulation replicates were prepared with the four terminus atoms G,C@O$_{5'}$,O$_{3'}$ subjected to constant-force restraints of strengths $(15,20..185,190)$ pN, oriented towards fixed sites at the top and bottom of the simulation box.  Each replicate was equilibrated in a box of 2446 TIP3P water molecules and 2 sodium ions  for 50ns.  During equilibration, frames evaluated with a low confidence (small denominator of eqn.~\ref{eqn:kernel}) were saved, and those points with the lowest confidence were used to augment the training set to 1232 frames. In production, force-extension time series were collected for 1500ns per replicate, with a 1fs timestep.   

The calculation was run with DNA bases restrained to Watson-Crick pairing, following standard B-DNA geometry \cite{Saenger1984} with the intention to therefore focus on the behaviour of stack interactions. In detail the restraints were such that there was no energy penalty for the proton-acceptor pairs G@N$_1$H-C@N$_3$, G@NH$_2$-C@O, C@H$_2$-G@O to be within distances $\pm$0.2\AA~of their equilibrium values, 1.94\AA, 1.85\AA, 2.95\AA. Beyond these ranges, harmonic restraints were imposed with a spring constant of 20 kcal mol$^{-1}$ \AA$^{-2}$, for a further 0.5\AA, after which restraints became constant-force.  For the two stronger hydrogen bonds, G@N$_1$-C@N$_3$ and C@N-G@O donor-acceptor distance restraints were also added with the same pattern, at equilibrium  distances 2.5\AA~and 2.51\AA, thus (weakly) enforcing planarity of the base pairs.\\

\subsection{Equilibrium Potential of Mean Force Calculations}

As an examination of the equilibrium behaviour of the KMMD calculation, potentials of Mean Force (PMF) over $\epsilon,\zeta,\chi$ dihedral angles were calculated using sets of multiple parallel simulations with and without KMMD.  Sampling was enhanced using the technique of Hamiltonian Replica Exchange (HREMD) within the sets of 120 simulations, which individually sampled umbrellas spaced 3 degrees apart in the torsion for which the given PMF was desired.  Umbrella torsion restraints were harmonic with the spring constant 250 kcal mol$^{-1}$ rad$^{-2}$.  The two chains were kept together using a weak flat-bottomed distance restraint between the two phosphorous atoms, with zero force at distances between 20\AA~and 30\AA~but a harmonic force with a spring constant of 20 kcal mol$^{-1}$ \AA$^{-2}$ thereafter.  Exchange flux between replicas was verified as leading to a physically credible free energy distribution, however the PMF were calculated using the Weighted Histogram Analysis Method (WHAM), implemented using the utility distributed by Alan Grossfield. \cite{Grossfield2002}

Replica exchange runs were chunked into sections of 10 ps (10000 steps of 1 fs each).  Each chunk had 2000 cycles of exchange attempts (one per 5 MD steps), with an average acceptance rate $r\approx$0.3.  The first chunk for each run was discarded as equilibration, and the subsequent two were used as production runs.   A Langevin thermostat with a temperature of 300 Kelvin and a coupling rate of 1 ps$^{-1}$ was applied to accelerate divergence between the 120 replicas in each set. Implicit solvent was applied using a Generalised Born solvation model. \cite{Nguyen2013} No cutoff was applied to the non-bonded interactions.  When errorbars were calculated with the assumption of a decorrelation time (per replica) equal to 100 fs, or 20 cycles of replica exchange, these appeared as almost too small to be visible.  Production runs for each equilibrium calculation comprised in total 120$\times$20 ps, or 2.4 ns, of dynamics with HREMD-accelerated sampling. 

\section{\label{sec:Results}Results}

\subsection{Correction versus Multiple Forcefields}

\begin{figure}[ht]
\begin{centering}
\includegraphics[width=1.0\columnwidth]{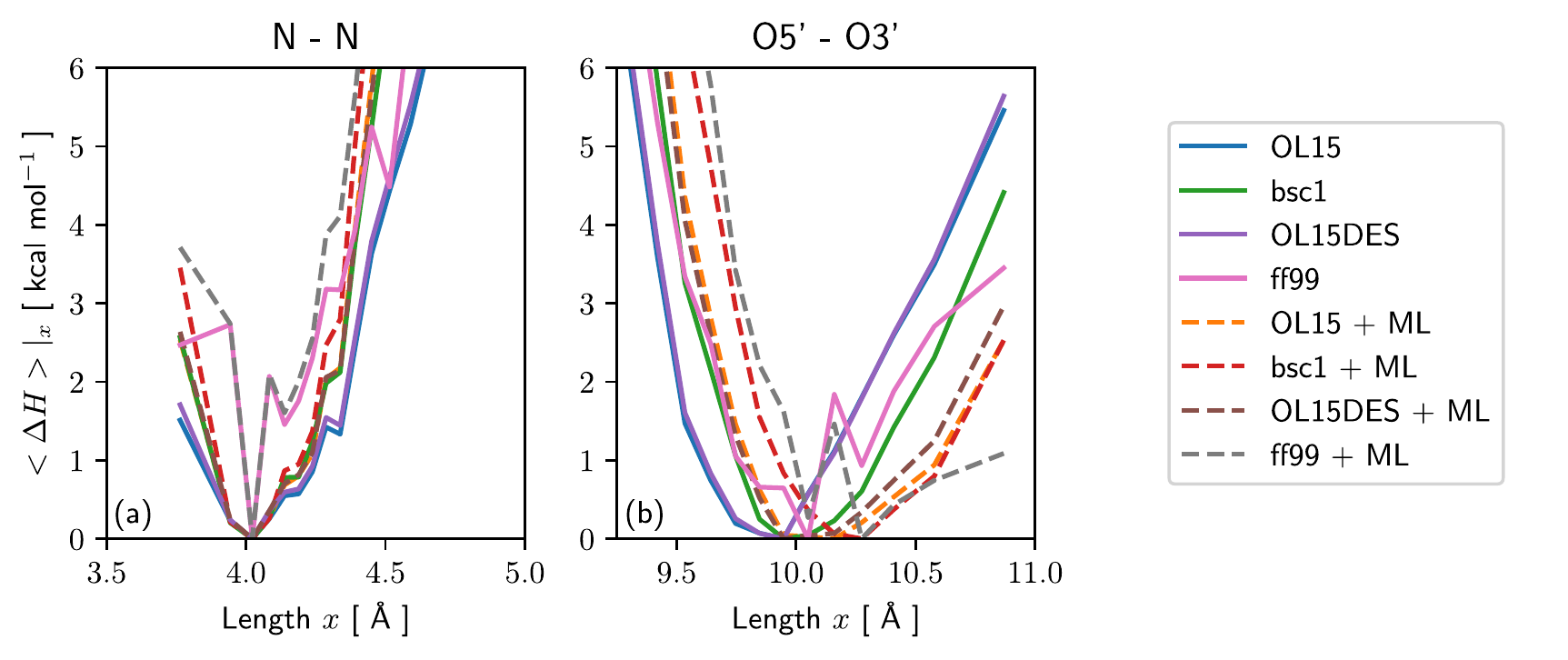}
\end{centering}
    \caption{\label{fig:cmpff}  Postprocessing frames with different forcefields shows that the differences between them are largely independent of the larger errors fixed by using the KMMD Machine Learning correction (marked as `ML'). \textbf{(a)}: Base-base distance (measured N9-N9 or N1-N1) shows moderate effects, with ML slightly increasing stiffness, but only when far from the minimum. \textbf{(b)}: Extension measured between the O5' and O3' termini is noticeably softer after correction, and has a shifted minimum, regardless of the starting
    forcefield, in a seeming conflict with the tendency of bases to resist having large separation.} 
\end{figure}

Stretching simulations were made against the OL15 forcefield, which is considered to be among the best classical DNA models at time of writing, \cite{Dans2017} however it was found prudent to make a survey of multiple forcefields in order to confirm that the deficiency in treating stacking is general (fig.~\ref{fig:cmpff}).  The bsc1 forcefield \cite{Ivani2016} was therefore compared against OL15 by post-processing snapshots taken from the KMMD simulations. The DESRES refinement of non-bonded interaction parameters for nucleobases \cite{Tan2018} was also checked, using the OL15 bonded and other non-specified parameters (`OL15DES').

Firstly snapshots were ordered by the length of the average vector between chain-adjacent bases measured at the N atoms where the chain joins the backbone (fig.~\ref{fig:cmpff}a),  in this case the behaviour around the minimum remained unchanged by adding KMMD, although the potential became apparently stiffer for large separation of the bases.  The minimum appeared very sharp with ff99, whether corrected or not, this effect arises because all frames were generated using OL15+KMMD and are therefore close to the energy minima or valleys as defined by that PES, other forcefields may have similar features but subject to slight geometry shifts, therefore giving apparently stiffer potentials.  When lengths $x$ were measured using the terminus atoms O5'-O3' (fig.~\ref{fig:cmpff}b) an increase in the equilibrium length of the duplex was observed.  This shift of $\approx 0.25$\AA~was independent of the base classical forcefield, and was associated with an continued softening of the potential with respect to further extension. All three forcefields performed similarly for snapshots having small end-to-end length $x$, at large $x$ the \textbf{bsc1} forcefield had the smallest over-estimate of pulling energy.  The differences between the forcefields seem to be largely independent of the KMMD correction: all were altered by the KMMD by quite similar amounts.

\subsection{Modulations of Collective Dynamics}

\begin{figure}[ht]
\begin{centering}
\includegraphics[width=1.0\columnwidth]{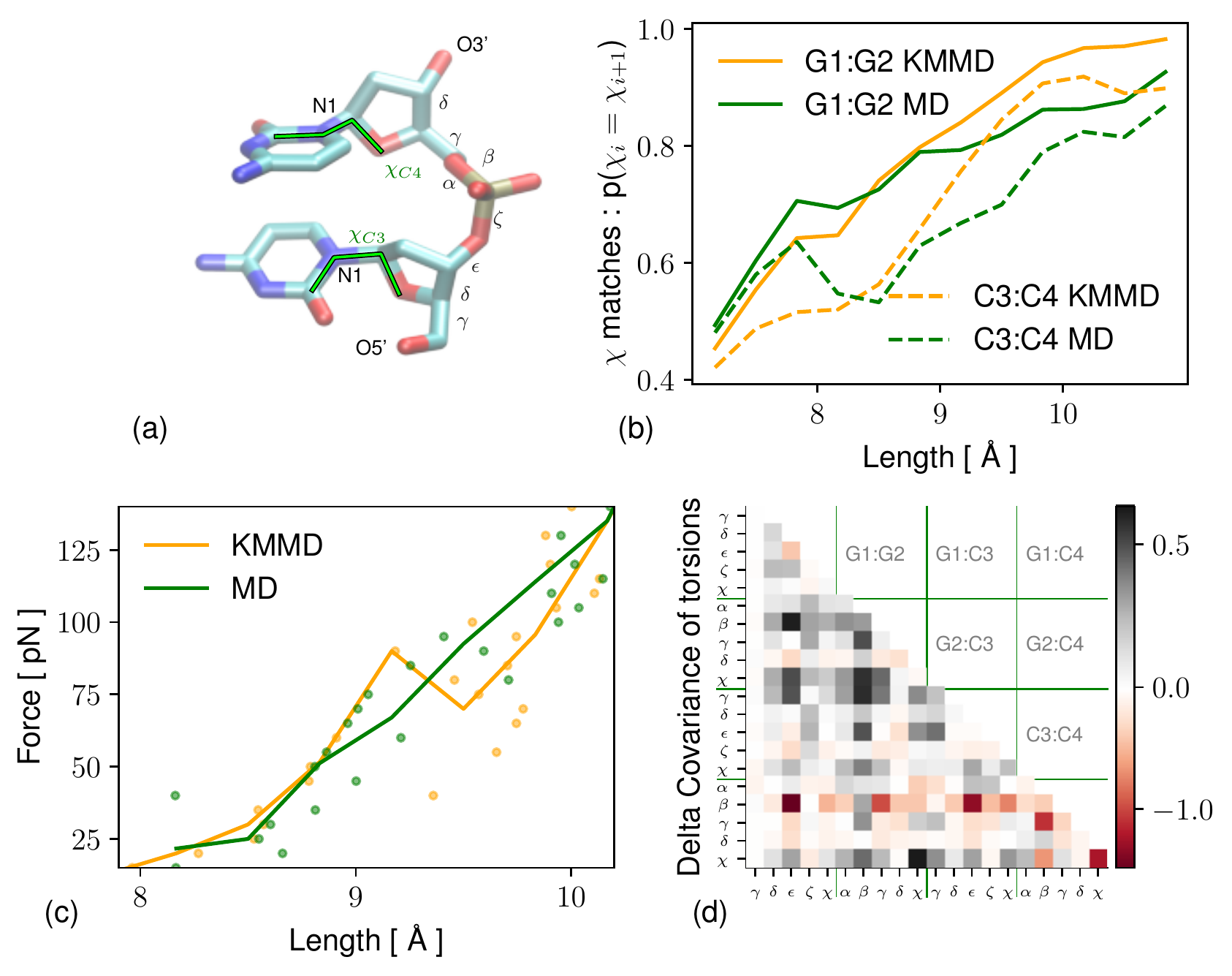}
\caption{\label{fig:chiExtWC} (a) The CC chain of the GG$\cdot$CC duplex, with backbone and $\chi$ torsions labelled.  The four atoms defining each $\chi$ are shown with green highlight lines. (b) Base-base interactions versus extension: proportion of $\chi$ angles on the same chain to be in the same basin.  At short extension, KMMD gives less correlation between adjacent $\chi$, at long extension KMMD strengthens the correlation.  (c) Force-extension. Points are independent constant-force simulations; lines are a moving average. (d): Change in covariance of dihedrals, KMMD-MD, over all extensions. Significant signals for $\chi$ and for torsions adjacent to central phosphate. }
\end{centering}
\end{figure}

The purpose of the KMMD correction to the forcefield is not primarily to moderate individual dihedral potentials, but to treat collective interactions in molecules with multiple soft torsions.  Fig.~\ref{fig:chiExtWC} compares structural correlations in the DNA duplex under stretch with and without the ML correction.  A basic probe of correlation is to collect the proportion of frames in which the $\chi$ angles adjacent on the chain were both in the same energy basin (of three available, as determined by the tetrahedral C1' atom at which the base joins the sugar).  As expected, the C3:C4 stacked pair of bases has an overall smaller chance for adjacent $\chi$ angles to be in the same minimum than does the G1:G2 stack, due to the smaller size of the C base.  For both stacks, there is a changeover in the effect of the KMMD with increasing extension: at short extension, KMMD destabilises the base stack relative to standard MD with the OL15 forcefield, while at higher extension, ordering of bases increases for all systems, but especially so with the KMMD (Fig.~\ref{fig:chiExtWC}b). This signal is consistent with stabilisation of edge-edge interactions by the KMMD when under imposed extension, as investigated in figures (\ref{fig:2dscans}, \ref{fig:mostFavoured}).

As simulations were run at constant force it is proper to show the force-extension curve (Fig.~\ref{fig:chiExtWC}c), although this observable is noisy and subject to finite size effects it does support a reduction of the work to extend the DNA base step, reducing the discrepancy against the lower work to force-melt macroscopic DNA typically seen in experiment relative to simulation. Although the KMMD force-extension trace appears visually to be flatter than the uncorrected trace, an outlier point at 90 pN has the effect of giving almost equal values for the work to extend from 8\AA~to 10\AA~when a simple integration is performed using the trapezium rule (1.69 kcal mol$^{-1}$ with KMMD versus 1.72 kcal mol$^{-1}$ without). If the offending datapoint is removed, the corrected work becomes 1.56 kcal mol$^{-1}$, giving a difference which is still quite small in absolute terms but significant as a fraction of the total work to stretch the base-pair step.  The experimental value given by Yakovchuk \textit{et al.} \cite{Yakovchuk2006} for the base pair stacking energy of GG$\cdot$CC at 300K is salt-dependent, giving 1.5$\pm$0.1 kcal mol$^{-1}$ at a salt concentration of 45mM, equivalent to the selected 2Na$^+$ in 2446 water.

Fig.~\ref{fig:chiExtWC}d gives a sketch of the pairwise torsion angle correlations most affected by the KMMD correction.  Covariance $C_{1,2}$ of a pair of dihedrals $\theta_1,\theta_2$ is defined here as $\overline{(\theta_1-\bar{\theta_1})(\theta_2-\bar{\theta_2})}$ in units radians squared (the means $\bar{\theta}_i$ are found as $\mathrm{arctan2}(\overline{\cos \theta_i},\overline{\sin\theta_i})$).  The (signed) differences in the absolute values of the covariances $|C_{KMMD}|-|C_{MD}|$ are shown in fig.~\ref{fig:chiExtWC}d as a heatmap.  Overall the KMMD correction promotes disorder (negative delta covariance, orange) however covariance between some torsions is (more weakly) increased (positive delta, grey), in particular involving $\chi$ and $\zeta$ angles.  The $\beta$ angle linking G bases becomes more covariant (grey) while the $\beta$ angle linking the C bases becomes much less covariant, and more disordered.

%\begin{figure}[ht]
%\begin{centering}
%         \includegraphics[width=1.0\columnwidth]{project_angles.pdf}
%\end{centering}
%    \caption{\label{fig:projAngles} Averages projected onto key angular degrees of freedom. The $\chi$ angle \textbf{(a)} is (overall) well handled without ML, however the sugar pucker and backbone $\epsilon, \zeta$ torsions \textbf{(b-d)} show substantial deviations.   } 
%\end{figure}

\begin{figure}[ht]
\begin{centering}
         \includegraphics[width=1.0\columnwidth]{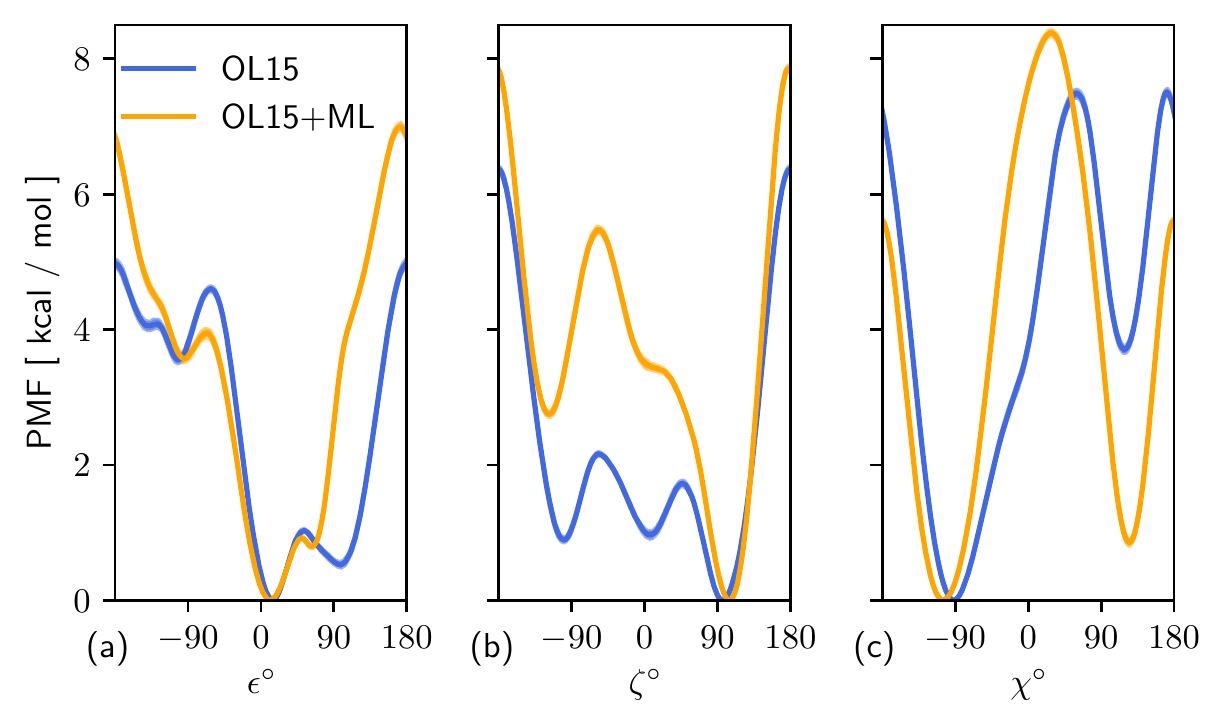}
\end{centering}
    \caption{\label{fig:projAngles} Potentials of mean force (PMF) for key dihedrals, sampled without any stretching or WC bonding imposed.} 
\end{figure}

\subsection{Conformational Analysis}

The strength of the ML method is that second-order and higher correlations between torsion angles are treated, however it is still instructive to project onto individual torsions in order to see which conformations of the duplex are stabilised/destabilised relative to classical forcefields.
% Barriers for rotation in $\zeta$ are lower using KMMD than otherwise, barriers in $\epsilon$ are higher, while the zero pucker angle (`North') is relatively disfavoured (fig.~\ref{fig:projAngles}). 
In order to compare the ML corrected forcefield with traditional nucleic acid forcefields, a series of replica exchange calculations were made (REMD) at zero imposed force and without hydrogen bond restraints, generating equilibrium potentials of mean force (PMFs).  Three of the most important dihedral angles were examined, the $\epsilon$ and $\zeta$ joining two G bases, and the $\chi$ angle of the first G. The qualitative picture was instructive: from the point of view of individual dihedral angles (as opposed to correlations between them) the KMMD correction is a step backwards through time.  Qualitatively, a 1-dimensional view of the KMMD correction seems to show it as operating in the opposite direction to a recently proposed correction to the AMBER ff10 forcefield,\cite{Aytenfisu2017} modifying the shapes of the free energy landscapes over the relevant dihedrals with the opposite sign relative to the original ff10 \cite{Perez2007,Zgarbova2011} (ff10 is a recent ancestor of the OL15 \cite{Zgarbova2015} starting point for the present study).  The correction of Aytenfisu \textit{et al.}~did not alter the functional form of the forcefield but did re-fit with a much wider range of nucleic acid construct structures, with the novelty being that many structures drawn from crystallographic data were used, as well as the typical systematic rotations of dihedral angles in otherwise canonical structures.  This use of crystallographic data selectively well-fits ordered structures. In the KMMD-corrected simulation, with training data drawn from highly disordered non-equilibrium structures, the less-dominant \textit{syn} conformation for the $\chi$ angle was relatively stabilised, individually promoting disorder in the base conformation even though it is clear from fig.~\ref{fig:chiExtWC} that collective order/disorder in base stacking without stretch is not much altered by KMMD. Training set bias is therefore less important when using KMMD because it is possible for a given dihedral to be both stabilised and destabilised at the same angle, depending on conformations of other dihedrals elsewhere in the system.  The 1D PMFs for $\epsilon$ and $\zeta$ appear, in constrast to $\chi$, to be order-stabilising in the KMMD without stretch, and this is consistent with the variance-covariance plot of fig.~\ref{fig:chiExtWC}d showing small or negative changes in the individual variances of these angles under stretch with the correction applied.  This is accentuated by a positive (gray) covariance between the adjacent $\epsilon$ and $\zeta$ angles: decreased individual variance coupled to increased covariance indicates an overall reduction of entropy associated to the backbone degrees of freedom without imposed stretch.  The behaviour of the $\chi-\chi$ couplings versus stretch in fig.~\ref{fig:chiExtWC}b indicate however a slight increase of base-stack entropy at low or no imposed stretch, so we can state that the classical forcefields have at equilibrium a slight under-stiffening of the backbone and a slight over-emphasis on base-base interactions or on the stiffness of the $\chi$ angles which also control base positioning.

\begin{figure}[ht]
\begin{centering}
         \includegraphics[width=1.05\columnwidth]{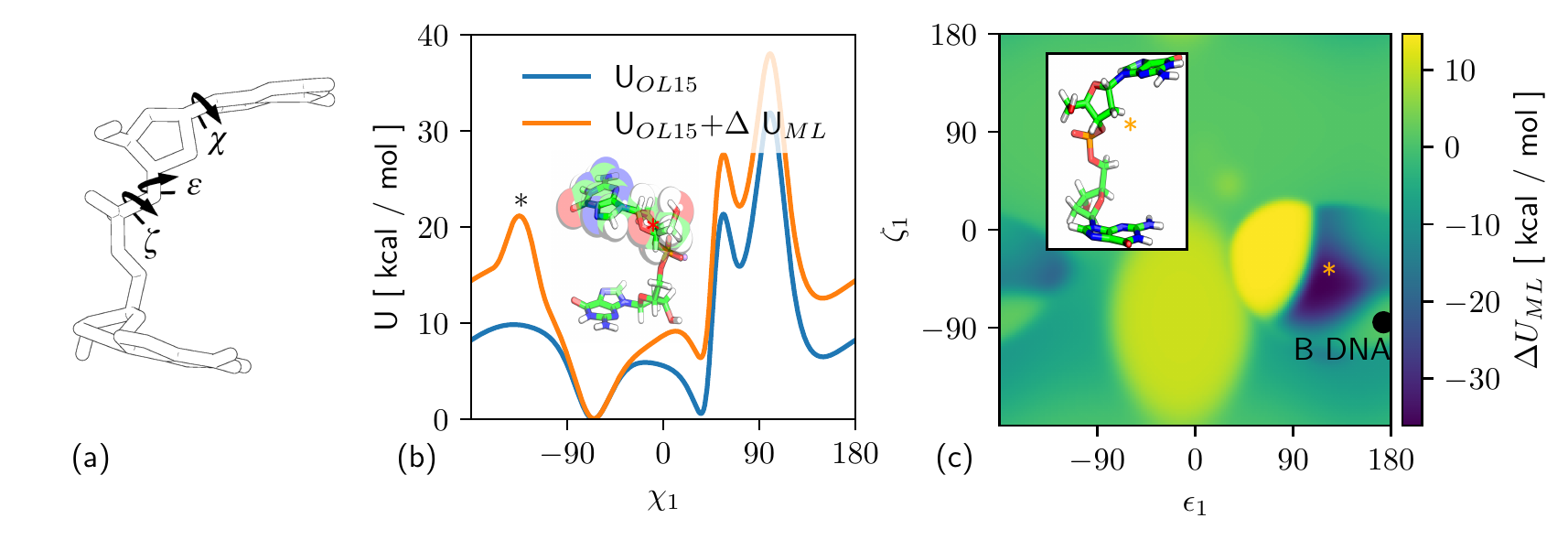}
\end{centering}
    \caption{\label{fig:2dscans} 1D and 2D cuts holding other angles fixed.  \textbf{(a)} Strand 1 (GG) of the dyad, showing angles permuted in the conformation examined.  \textbf{(b)} A steric clash (red and white spheres, \textit{inset foreground}) at $\chi=-140^{\circ}$ is penalised more harshly by the ML method.   \textbf{(c)} A highly extended conformation (* and inset) is strongly favoured by the ML. } 
\end{figure}

These data do not imply a simple revision of the existing backbone torsion potentials, and a re-fitting of individual dihedral angle potentials such that they could agree with the KMMD 1D-PMFs  would land close to the AMBER ff10, which is after more than a decade now regarded as relatively obsolete even among traditional forcefields.  What is shown however is that conflicting sets of training data (the dihedral scans which generated ff10 and the crystallographic structures which contributed to the Aytenfisu \textit{et al}.~ correction) can be reconciled by considering a larger number of degrees of freedom together.  Collective dihedral angle corrections can indirectly address errors arising from the other terms of the forcefield: the example of the steric clash between the hydrogen attached to the Guanine C8 (or to any purine C8 Hydrogen) and the pentose O4' is investigated by permuting the angles labelled in fig.~\ref{fig:2dscans}(a).  Although the Pauli exclusion which drives the sharp increase of energy as atoms approach overlap is not addressed directly by the KMMD correction; because molecular geometry is determined predominantly by dihedral angles, it is possible to indirectly address a clash which the standard classical treatment misses.

%support the understanding that sugar pucker and backbone torsions have non-trivial correlations which a conventional few-body forcefield might struggle to reflect. %% \cite{Neidle2008}.

\subsection{Conformations of Greatest Deviation}

\begin{figure}[ht]
\begin{centering}
         \includegraphics[width=0.5\columnwidth]{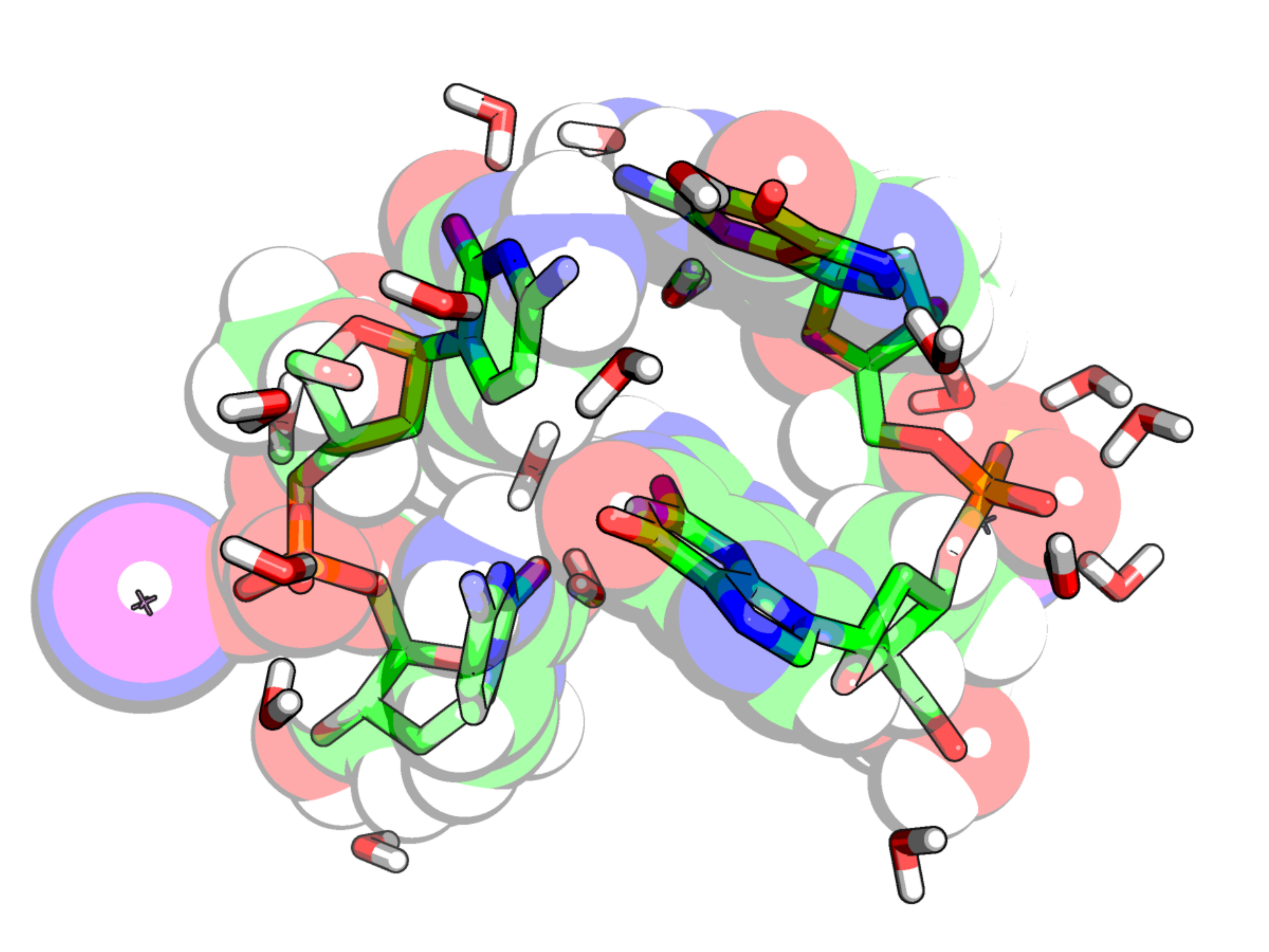}
\end{centering}
    \caption{\label{fig:mostFavoured} A conformation highly favoured by the ML relative to the uncorrected OL15 forcefield. Hydrogen bonding is preserved and bases on the same strand remain (sometimes/partly) in contact despite a large vertical stretch.} 
\end{figure}

Analysing the individual conformations with the greatest differences between KMMD and classical treatment interrogates the limitations of classical methods and helps to motivate the operation of the method.  The top 100 structures with large favourable corrections following KMMD are dominated by conformations showing roughly 90$^\circ$ base-base interactions (the single most-favoured conformation is shown in fig.~\ref{fig:mostFavoured}).  

We refer to stretched duplex geometries with mostly-preserved hydrogen bonding and partly-preserved stacking (subject to tilt and slide) as `$\tau$-steps' \cite{Taghavi2017}, these have been put forward among various candidate geometries for stretched DNA since the earliest molecular mechanics studies \cite{Lebrun1996}, and linked specifically to stretched GG$\cdot$CC pairs in recent classical MD work \cite{Shepherd2020}; our primary structural observation into DNA under tension therefore is that the posited $\tau$-step conformation is confirmed, and that this is very much more favourable than found when modelled classically.  The strength of this interaction geometry is not inconsistent with the chemical understanding of 90$^\circ$ interactions between aromatic molecules with hydrogen bonding groups at the edges, in general discussed as `remarkably strong'   \cite{Rutledge2008}, although the canonical T-shape aromatic geometry is typically presented as edge-face rather than edge-edge. The lack of directionality imposed by a pointlike treatment of classical atoms leaves little hope of correctly treating the highly directional combination of hydrogen bonding and aromatic stacking in the framework of current mainstream forcefields. 

\subsection{Outlook}

Distillation of \textit{ab initio} quantum mechanical calculations on fragments into a  smooth potential energy surface was successful in this initial example, demonstrating integration of machine-learned physics for short-range interactions with classical modelling of longer-range Coulomb and pairwise dispersion physics, as well as classical treatment of solvent and of stiffer short-range degrees of freedom (bond lengths and angles) not strongly affecting the geometry.  Generalisation to long chains as a sum of overlapping fragments does not seem in principle to be a major challenge, nor does generalisation to a wider array of treated repeating units such as amino acids or RNA.  Classical MD simulations corrected with KMMD were able to provide a qualitatively correct alteration to the force-extension behaviour of a small DNA construct, bringing results more closely into line with experiment.  KMMD does not as applied here treat chemical effects, collective electron dynamics in the solvent, or convergence errors due to rare fluctuations; however by application of Occam's razor following the results presented here we may reduce our concern in relation to these potential sources of error.

The KMMD concept as described here does ultimately have well-defined limitations, arising in particular from the fact that non-bonded interactions are treated indirectly only.  It is entirely feasible to construct an augmented feature space containing two-body or N-body distance information as well as the angular information treated in the present implementation, indeed this has been done for kernel methods targeting smaller molecules,\cite{Hansen2013} however there are always costs for adding complexity.  Development of accuracy and generality for kernel-modified MD is likely to continue in the near future by considered design of the feature space in which the kernel operates, and the thoughtful integration of physics which can be calculated cheaply and well from first principles with physics which is best treated through machine learning. 

KMMD was shown to modify DNA dynamics with a qualitatively correct reduction of the force to extend a small two-base pair construct in water, of the correct sign such that it can now be hoped that simulations of larger systems will manifest thermodynamics (and perhaps structure) in line with experiment. The putative biologically relevant conformations for many-base pair DNA which have been suggested from experimental data in relation to stretching \cite{Taghavi2017,Bosaeus2017} and also to hydrophobic denaturants \cite{Feng2019} were not susceptible to be confirmed or denied by this study, in particular the hypothetical $\Sigma$ phase of DNA in which symmetry is broken with a three-bp periodicity is obviously not susceptible to direct study by simulations of 2bp only.

%%%%%%%%%%%%%%%%%%%%%%%%%%%%%%%%%%%%%%%%%%%%%%%%%

\begin{acknowledgments}
The experiments presented in this paper were carried
out using the HPC facilities of the University of Luxembourg.\cite{hpcUniLu} Supported by the Luxembourg National Research Fund (FNR) grant C20/MS/14769845/BroadApp.
\end{acknowledgments}

\section*{Data Availability Statement}

Software to carry out the KMMD and other calculations is available from \url{ambermd.org} as a part of the AmberTools 22 package.\cite{Case2022} The dataset of quantum energy evaluations generated for this work is available from the NOMAD database under a Document Object Identifier: \url{https://dx.doi.org/10.17172/NOMAD/2022.04.25-1 10.17172/NOMAD/2022.04.25-1}.

%\nocite{*} %%nocite brings in *everything* in the bib file, cited or not.
\bibliography{trips.bib}
\end{document}